\documentclass{PoS}

\usepackage{graphicx}
\usepackage{amsmath}

\newcommand{\beq}{\begin{equation}}
\newcommand{\eeq}{\end{equation}}
\newcommand{\beqa}{\begin{eqnarray}}
\newcommand{\eeqa}{\end{eqnarray}}

\title{Nuclear effective field theory on the lattice}

\ShortTitle{Nuclear effective field theory on the lattice}

\author{\speaker{Hermann~Krebs}$^{a,c}$\\
E-mail:\,\email{h.krebs@fz-juelich.de}}
\author{Bu\={g}ra~Borasoy$^{a}$\\
E-mail:\,\email{borasoy@itkp.uni-bonn.de}}
\author{Evgeny Epelbaum$^{c,a}$\\
E-mail:\,\email{e.epelbaum@fz-juelich.de}}
\author{Dean~Lee$^{d}$\\
E-mail:\,\email{djlee3@ncsu.edu}}
\author{Ulf-G.~Mei{\ss }ner$^{a,b,c}$\\
E-mail:\,\email{u.meissner@fz-juelich.de}\\
\\
$^{a}$Helmholtz-Institut f\"{u}r Strahlen- und Kernphysik (Theorie)
Universit\"{a}t Bonn, \\
Nu\ss allee 14-16, D-53115 Bonn, Germany \\
\\
$^b$Bethe Center for Theoretical Physics,
  Universit\"at Bonn,\\ D-53115 Bonn, Germany\\
\\
$^{c}$Institut f\"{u}r Kernphysik (IKP-3) and J\"ulich Center for Hadron
Physics, \\
D-52425 J\"{u}lich, Germany \\
\\
$^{d}$Department of Physics, North Carolina State University, Raleigh,
NC 27695, USA
}

\abstract{In the low-energy region far below the chiral symmetry breaking scale of
  $\Lambda_\chi\sim 1$~GeV chiral perturbation theory (ChPT) provides
 a model-independent approach for quantitative description of nuclear
processes. 
In the two- and more-nucleon sector perturbation theory is applicable only at the
level of an effective potential which serves as input in the corresponding
dynamical equation. To deal with the resulting many-body problem we put
chiral effective field theory (EFT) on the lattice. 
Here we present the results of our lattice EFT study up to next-to-next-to-leading
order (N$^2$LO) in the chiral expansion. Accurate description of two-nucleon phase-shifts  and ground state energy 
ratio of dilute neutron matter up to corrections of higher 
orders shows that lattice EFT is a promising tool for a
quantitative description of low-energy few- and many-body systems.
}

\FullConference{The XXVI International Symposium on Lattice Field Theory\\
		 July 14-19 2008\\
		 Williamsburg, Virginia, USA}

\begin{document}
\section{Introduction}
Quantum chromodynamics (QCD) describes the interaction between quarks and 
gluons which is responsible for the strong nuclear force. Recent advances in 
QCD using
computational lattice methods have made it possible to accurately predict the
spectrum and properties of many isolated hadrons. Unfortunately, lattice QCD
calculations of nuclear and neutron matter or even
few-body systems beyond two nucleons are presently not possible. 
The most
significant challenge in such simulations would be to overcome the 
exponentially small signal-to-noise ratio caused by the sign and complex phase 
oscillations for simulations at large quark number.

Nuclear lattice simulations based on EFT provide an 
alternative method to describe few- and many-body systems at low energy
 without losing connection to QCD. The lattice EFT
approach addresses the few- and many-body problem in nuclear physics by
applying non-perturbative lattice methods to low-energy nucleons and pions.
The effective Lagrangian is formulated on a spacetime lattice and the path
integral is evaluated by Monte Carlo sampling. Pions and nucleons are treated
as point-like particles on the lattice sites. By using hadronic degrees of freedom and concentrating on low-energy physics, 
it is possible to probe large
volumes and greater number of nucleons than in lattice QCD. After a brief
overview what has been done in this field so far 
we 
present some results of our recent studies of the two nucleon 
system~\cite{Borasoy:2007vy}\cite{Borasoy:2007vi} and 
neutron matter~\cite{Borasoy:2007vk} at subleading order. 
Accurate description of two-nucleon phase-shifts  and ground state energy 
ratio of dilute neutron matter up to corrections of higher 
orders show that lattice EFT is a promising tool for
quantitative studies of low-energy few- and many-body systems.

\section{Lattice EFT: previous achievements}

Lattice EFT is a rather new and fast developing field.
Here we give a brief overview on what has been done in this field so far. 
For a comprehensive discussion the reader is referred to~\cite{Lee:2008fa}.
The first lattice study of nuclear matter was carried out in the early nineties
by Brockman and Frank~\cite{Brockmann:1992in} using a momentum 
lattice and based on the hadrodynamics model of
Walecka~\cite{Walecka:1974qa}. 
The first nuclear lattice calculation based on EFT was carried out by M\"uller
et al.~\cite{Muller:1999cp}. They looked at infinite nuclear and neutron 
matter at nonzero density and temperature. Later a series of 
analytical studies were carried out: Chen and Kaplan~\cite{Chen:2003vy}
 showed the absence of
sign oscillation for nonzero chemical potential in the Hubbard model. Non-linear
realization of chiral symmetry with static nucleons on the lattice was 
discussed by Chandrasekharan et al.~\cite{Chandrasekharan:2003ub}. Also ChPT
 within the lattice regularization was considered by
several groups~\cite{Shushpanov:1998ms,Lewis:2000cc,Borasoy:2003pg}. This was 
followed by the first many-body lattice calculation using chiral
EFT~\cite{Lee:2004si}. Since that time 
a number of lattice calculations for cold atoms and low-energy nuclear
physics were carried out. See~\cite{Lee:2008fa} for a review article. 
It is important to note that in the 
low-energy sector the phase region accessible by lattice EFT
 is much broader than in lattice QCD. Severe sign
oscillation problem limits the accessibility of finite
density lattice QCD simulations. In contrast, sign oscillations
in nuclear lattice EFT are strongly suppressed due to the
approximate $\rm{SU}(4)$-symmetry in the two-nucleon sector. One can 
show explicitely that $\rm{SU}(4)$-symmetric nuclear EFT
does not have a sign problem. $\rm{SU}(4)$ symmetry breaking leads
to small sign oscillations which, however, turns out to be not severe.

\section{Nuclear EFT}
Let us now give a brief introduction to the basic foundations of our approach.
The low-energy properties of hadronic systems are, in principle, accessible
in lattice QCD. This method is, however, very expensive, especially if one
wants to consider few- and many-nucleon systems.
 Alternatively, we can exploit the spontaneously broken approximate chiral symmetry of QCD which implies the
existence of light weakly interacting Goldstone bosons. 
In the ${\rm SU}(2)$ sector, we identify the Goldstone bosons with pions. Since
the interaction between the Goldstone bosons is weak one can apply 
perturbation theory, where the expansion
parameter is not a coupling constant but small momenta and masses of the
Goldstone bosons divided by the chiral symmetry breaking scale
$\Lambda_\chi$. 
This systematic
procedure is called chiral perturbation theory~\cite{Gasser:1983yg} 
and reproduces (as explicitely proved by 
Leutwyler~\cite{Leutwyler:1993iq} in mesonic sector) order by
order the original QCD Green-functions.

ChPT has been extended to one nucleon sector. In the
two and more nucleon sector additional problems appear. Due to the existence of
nuclear bound states, the strict perturbative procedure breaks down.
As shown by Weinberg~\cite{Weinberg:1991um}, the power counting
is violated by nucleon-nucleon (NN) cuts. 
He suggested to construct perturbatively
a so-called chiral effective potential which, per construction,
excludes the NN 
cuts and, for this reason does not violate the power counting of ChPT. 
To describe nuclear observables in the two-, three- or
more-nucleon sectors one should numerically solve the
Lippmann-Schwinger, Faddeev or Faddeev-Jakubowsky equations, respectively,
 with the chiral effective potential as an input. 

\begin{figure}[t]
\includegraphics[width=15.0cm,keepaspectratio,angle=0,clip]{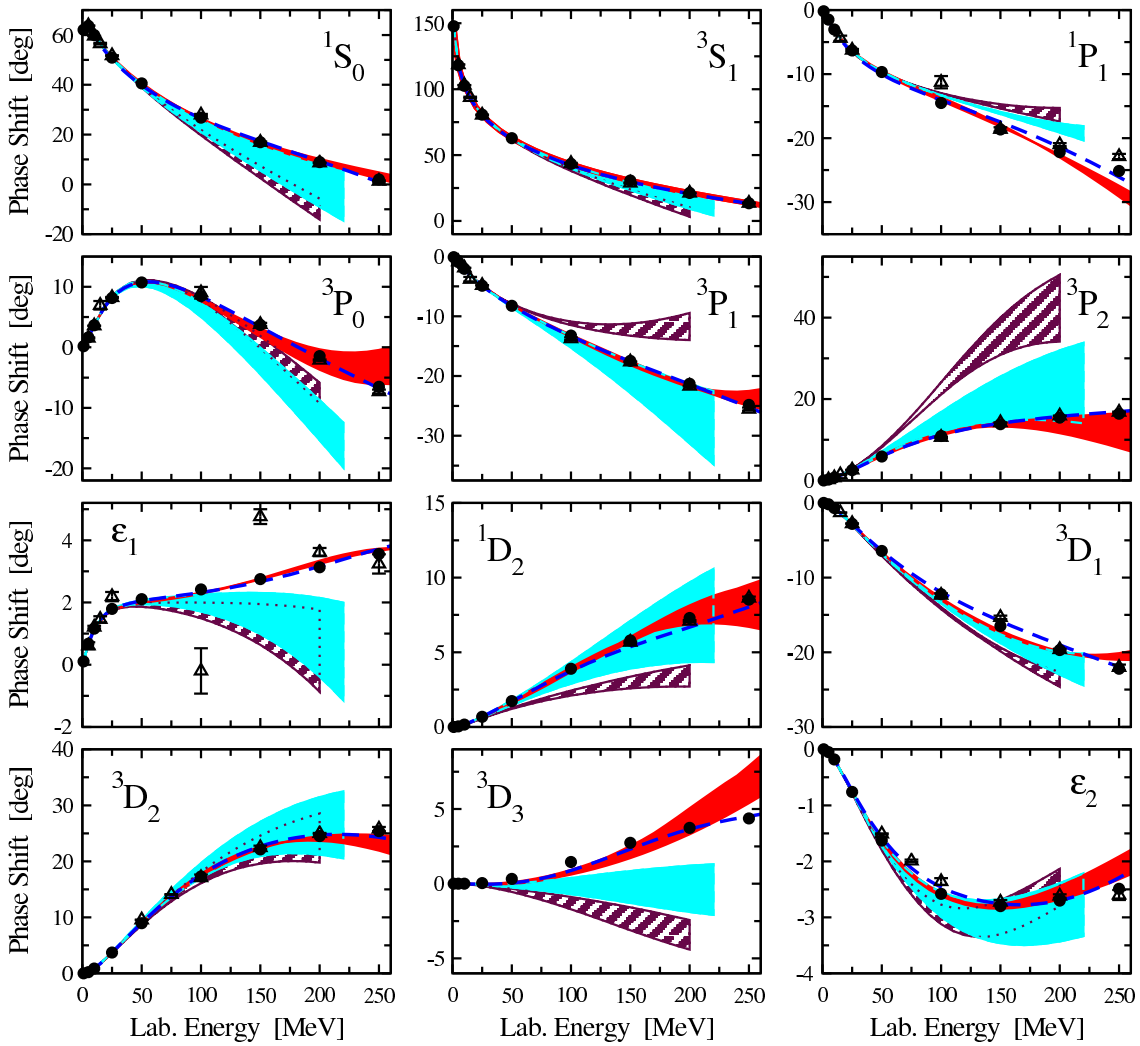}
\vspace{-0.15cm}
\caption[fig23]{\label{fig23} S--, P-- and D--waves {\it np} phase shifts. 
The dashed, light shaded and dark shaded bands show the NLO, N$^2$LO and N$^3$LO  
\cite{Epelbaum:2004fk} results, respectively. 
The dashed line is the N$^3$LO result of Ref.~\cite{Entem:2003ft}. 
The filled circles (open triangles) depict the results from the Nijmegen multi--energy PWA \cite{Stoks:1993tb,NNonline} 
(Virginia Tech single--energy PWA \cite{SAID}).  }
\vspace{0.2cm}
\end{figure}

\begin{figure}[t]
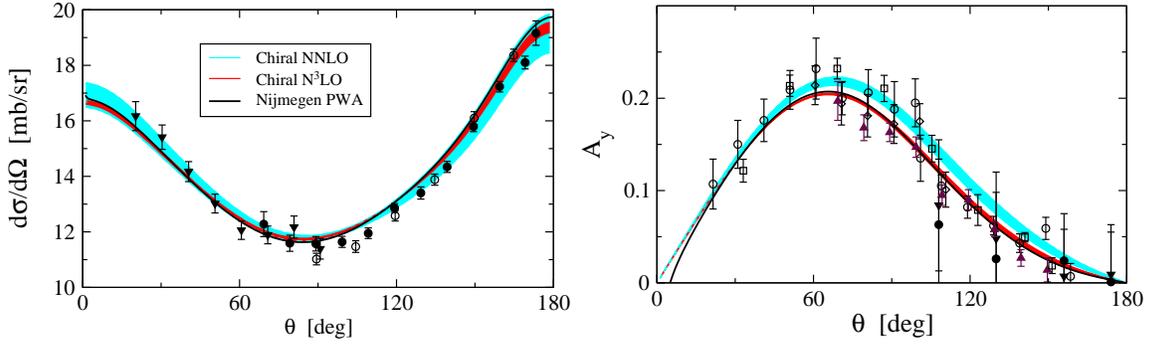

\includegraphics[width=7.5cm,keepaspectratio,angle=0,clip]{ds50sw.eps}
\includegraphics[width=7.5cm,keepaspectratio,angle=0,clip]{ay50sw.eps}
\vspace{-0.15cm}
\caption[fig25]{\label{fig25}{\it np} differential cross section and vector
  analyzing power at $E_{\rm lab} = 50$ MeV. The Nijmegen PWA result is taken
  from \cite{NNonline}.}
\vspace{0.2cm}
\end{figure}

Chiral effective potential has been extensively studied in the last decade
up to next-to-next-to-next-to-leading order~(N$^3$LO) in chiral 
expansion (for extensive discussion see~\cite{Epelbaum:2005pn}). At this order 
two leading-order (LO), seven subleading order (NLO) and fifteen N$^3$LO unknown low energy constants  
have been fitted to low energy nucleon
data~\cite{Epelbaum:2004fk,Entem:2003ft,Stoks:1993tb}. 
At this order in the chiral expansion, one observes an accurate description of
all NN low-energy observables, see Figs.~\ref{fig23},~\ref{fig25} and 
Table~\ref{tab:D}.

\begin{table*}[t] 
\begin{center}
\begin{tabular*}{1.0\textwidth}{@{\extracolsep{\fill}}||l||c|c|c||c||}
    \hline \hline
{} & {} &  {} &  {} & {}\\[-1.5ex]
    & {NLO}    & {N$^2$LO}  & {N$^3$LO}   &  {Exp} \\[1ex]
\hline  \hline
{} & {} &  {} &  {} & {}\\[-1.5ex]
$E_{\rm d}$ [MeV] &   $-2.171 \ldots  -2.186$   & $-2.189 \ldots -2.202$   & $-2.216 \ldots -2.223$  & $-$2.224575(9) \\[0.3ex]
$\eta_{\rm d}$ &     $0.0256 \ldots   0.0257$     & $0.0255 \ldots 0.0256$   & $0.0254 \ldots 0.0255$    & 0.0256(4)\\[0.3ex]
%
$A_S$ [fm$^{-1/2}$] & $0.868\ldots 0.873$ & $0.874  \ldots  0.879$   &  $0.882 \ldots 0.883$    & 0.8846(9)\\[0.3ex]
    \hline \hline
  \end{tabular*}
\caption{Deuteron properties at NLO, N$^2$LO and N$^3$LO compared to the data. Here, $E_{\rm d}$ is the
    binding energy, $\eta_{\rm d}$ the asymptotic
    $D/S$ ratio and  $A_S$ the 
    strength of the asymptotic S--wave normalization. The data for $E_{\rm d}$ are from \cite{Leun:1982aa},
    for $\eta_{\rm d}$ from \cite{Rodning:1990zz} and for $A_S$ from \cite{Ericson:1982ei}.  
    \label{tab:D}}
\end{center}
\end{table*}
\section{Nuclear EFT on the lattice}
Once the chiral nuclear forces are determined and the low energy constants
appearing in the nuclear forces are fitted (in the two and three-nucleon
sector) one can make predictions in the four- and more-nucleon sectors
based on chiral EFT.
However, explicit numerical 
treatment of the Jakubowsky equations for more than four nucleons is a very
difficult task. To solve the many-body problem we propose to put 
the chiral effective potential on the lattice and apply the powerful 
Monte-Carlo
techniques which are already developed to high degree. In this framework,
nucleons are represented as point-like Grassman-fields and pions as point-like 
instantaneous~(in order to reproduce the chiral potential) pseudoscalar 
fields. Typically, our calculations are carried out using the lattice
length $L\simeq 20$~fm and the lattice spacing $a\simeq 2$~fm which corresponds
to the cutoff $\Lambda=\pi/a\simeq 300 \,{\rm MeV}$.
The correlation
function  for $A$ nucleons in the Euclidean space is defined by
\beq
Z_A(t)=\langle\Psi_A|\exp(-t H)|\Psi_A\rangle,
\eeq
where the states $|\Psi_A\rangle$ refer to the slater determinants for $A$ free 
nucleons, $H$ is the Hamiltonian of the system and $t$ the Euclidean time. The 
ground state energy of the $A$-nucleon system can be derived from the 
asymptotic behavior of the correlation function for large $t$.
\beq
E_A^0=-\lim_{t\rightarrow\infty}\frac{d}{d t}\ln Z_A(t).
\eeq
Expectation value of a normal ordered operator ${\cal O}$ can be derived in
a similar way by
\beq
\langle\Psi_A^0|{\cal O}|\Psi_A^0\rangle=\lim_{t\leftarrow\infty}
\frac{Z_A^{{\cal O}}(t)}{Z_A(t)}, \quad
Z_A^{{\cal O}}(t)=\langle\Psi_A|\exp(-t H/2){\cal O}\exp(-t H/2)|\Psi_A>,
\eeq
where the states $|\Psi_A^0\rangle$ denote the ground states of 
$A$-nucleons system. It is convenient to describe NN contact
interactions by standard bilinear nucleon density operators using the
Hubbard-Stratonovich transformation. Using the relation
\beq
\exp(\rho^2/2)\sim\int ds\, \exp(-s^2/2-s \,\rho)
\eeq
 one can
express terms quadratic in the nucleon density operator $\rho$ as terms 
linear in $\rho$ in the presence of auxiliary background fields. In this
representation, the full correlation function is related to the path integral
over pions and auxiliary fields,
\beq
Z_A(t)\sim\int \prod_{I=1,2,3}D\pi_I\prod_iDs_i\exp(-S_{\pi\pi}-S_{ss})
\langle\Psi_A|M^{(L_t-1)}(\pi_I,s_i)\cdots M^{(0)}(\pi_I,s_i)|\Psi_A\rangle.
\eeq 
Here $S_{\pi\pi}$ and $S_{ss}$ are free actions for pions and auxiliary fields
$s_i$ and $M^{(n)}$ is a transfer matrix defined as an $n$'th step in the
temporal direction. Note since we only have linear nucleon density
operators in the action the amplitude
\beq
\langle\Psi_A|M^{(L_t-1)}(\pi_I,s_i)\cdots M^{(0)}(\pi_I,s_i)|\Psi_A\rangle
\eeq
is just a slater determinant of single nucleon matrix elements 
${\cal M}_{i,j}$ with $i,j=1,\dots,A$. 
\section{Lattice EFT at leading order}
To be specific, we give here the leading order action starting with the free
theory. The presentation here is somewhat sketchy. For an extensive discussion 
see~\cite{Borasoy:2006qn}. The free auxiliary fields and pion actions are given by
\beq
S_{ss}(s,s_I)=\frac{1}{2}\sum_{\vec{n}}s(\vec{n})^2
+\frac{1}{2}\sum_{I=1}^3\sum_{\vec{n}}s_I(\vec{n})^2,\quad
S_{\pi\pi}(\pi_I)=\frac{\alpha_t}{2}\sum_{I=1}^3\sum_{\vec{n}}
\pi_I(\vec{n})(-\Delta+M_\pi^2)\pi_I(\vec{n}),
\eeq
where $M_\pi$ is the physical pion mass, $I$ denotes isospin indices and 
$\alpha_t=a_t/a$. For nucleons we use
$O(a^4)$ improved free lattice Hamiltonian defined by
\beq
H_{{\rm free}}=\frac{1}{m}\sum_{k=0}^3\sum_{\vec{n}_s,\hat{l}_s,i,j}f_k
\left[a_{i,j}^\dagger(\vec{n}_s)\left(a_{i,j}(\vec{n}_s+k \hat{l}_s)
+ a_{i,j}(\vec{n}_s-k \hat{l}_s)\right)\right], 
\eeq
where $m$ is the nucleon mass, the operators $a_{i,j}^\dagger(\vec{n}_s)$ and 
$a_{i,j}(\vec{n}_s)$ are the
nucleon creation and
annihilation operators, $\vec{n}_s$ are spatial coordinates, $\hat{l}_s$ 
are spatial unit vectors,
the indices $i$ and $j$ stay for spin and isospin indices, respectively, and
the coefficients $f_k$ read:
\beq
f_{0,1,2,3}=\frac{49}{2},-\frac{3}{4},\frac{3}{40},-\frac{1}{180}.
\eeq
To define the interactions we introduce the nucleon-density operators with
different spin/isospin polarizations
\beq
\rho^{a^\dagger,a}(\vec{n}_s)=\sum_{i,j}a_{i,j}^\dagger(\vec{n}_s)
a_{i,j}(\vec{n}_s), \quad
\rho_I^{a^\dagger,a}(\vec{n}_s)=\sum_{i,j,j^\prime}
a_{i,j^\prime}^\dagger(\vec{n}_s)[\tau_I]_{j^\prime,j}a_{i,j}(\vec{n}_s),
\eeq
\beq
\rho_{I,S}^{a^\dagger,a}(\vec{n}_s)=\sum_{i,i^\prime,j,j^\prime}
a_{i^\prime,j^\prime}^\dagger(\vec{n}_s)[\sigma_S]_{i^\prime,i}[\tau_I]_{j^\prime,j}a_{i,j}(\vec{n}_s).
\eeq
The transfer matrix for $n_t$-th step has, besides the free part, two important 
contributions:
\beqa
M^{(n_t)}&=&:\exp\left\{-H_{{\rm free}}\alpha_t
-\frac{g_A\alpha_t}{2 F_\pi}\sum_{S,I}\sum_{\vec{n}_s}\nabla_S
  \pi_I(\vec{n}_s,n_t)\rho_{S,I}^{a^\dagger,a}(\vec{n}_s)\right.\nonumber\\
&+&\left.\sqrt{-C\alpha_t}\sum_{\vec{n}_s}\left[s(\vec{n}_s,n_t)\rho^{a^\dagger,a}(\vec{n}_s)
+i\sqrt{C_I\alpha_t}\sum_{I}s_I(\vec{n}_s,n_t)\rho_I^{a^\dagger,a}(\vec{n}_s)\right]\right\}:.
\label{trasferMLO1}
\eeqa
Here $::$ denotes normal ordering. The first long-range contribution
includes the instantaneous pion-nucleon interaction and describes the
one-pion-exchange in the leading-order effective potential. The second
short-range contribution corresponds to the NN contact interactions. The
low-energy constants $C$ and $C_I$ fitted to Nijmegen PWA appear to
have different signs:
\beq
C<0,\quad C_I>0.
\eeq
With these signs the pion-less theory appears to have no sign-oscillations if
the number of protons and neutrons are equal and they stay pair-wise in
isospin-singlet states. In this case the multiplication with $\tau_2$ of the single-nucleon
matrix elements ${\cal M}$ from left and right is well defined and gives
\beq
\tau_2{\cal M}\tau_2={\cal M}^*.
\eeq
For this reason, the determinant of ${\cal M}$ appears to be real:
\beq
\det{\cal M}^*=\det{\cal M}.
\eeq
Since $\tau_2$ is antisymmetric, the eigenvalues of ${\cal M}$ are doubly
degenerate. This leads to a positive slater determinant~\cite{Chen:2003vy,Lee:2004ze}
\beq
\det{\cal M}\ge 0.
\eeq
The introduction of pions causes small sign-oscillations which, however, are
not severe and appear to be suppressed.

\begin{figure}[t]
\begin{center}
\includegraphics[width=12.0cm,keepaspectratio,angle=0,clip]{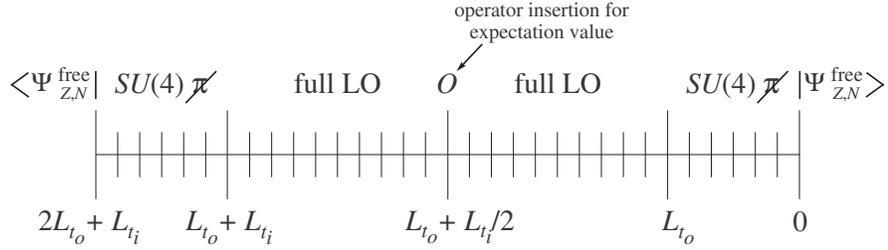}
\vspace{-0.15cm}
\caption[timesteps]{\label{timesteps} Overview of the various pieces of the
  transfer matrix calculation.}
\end{center}
\vspace{0.2cm}
\end{figure}

To perform our simulations in a most efficient way we, in addition, exploit the
approximate ${\rm SU}(4)$-Wigner~\cite{Wigner:1937zz} symmetry in the
NN system. The symmetry transformation is given by independent
rotation of spin and isospin degrees of freedom.
\beq
\delta N=\alpha_{\mu\nu}\sigma^\mu\tau^\nu N \quad {\rm
  with} \quad\sigma^\mu=(1,\vec{\sigma})\quad{\rm and}\quad \tau^\mu=(1,\vec{\tau}).
\eeq
One can show that in the limit where the NN S-wave scattering
lengths approach infinity the two-nucleon system becomes invariant under the
${\rm SU}(4)$-transformation~\cite{Mehen:1999qs}. The ${\rm SU}(4)$-breaking corrections come from the finite
scattering length and higher order terms in the chiral expansion:
\beq
{\rm SU}(4)-{\rm breaking\,terms}\sim\frac{1}{a(^3S_1)}-\frac{1}{a(^1S_0)},\frac{q}{\Lambda_\chi}.
\eeq
Since the NN scattering lengths
\beq
a(^1S_0)=(-23.758\pm 0.010){\rm fm}\quad a(^3S_1)=(5.424\pm 0.004){\rm fm}
\eeq
are very large, the ${\rm SU}(4)$-breaking corrections appear to be small. This
fact can be used to improve the performance of our lattice simulations. The
${\rm SU}(4)$ symmetric transfer matrix is given by
\beq
M^{(n_t)}=:\exp\left[-H_{{\rm
      free}}\alpha_t+\sqrt{-C\alpha_t}\sum_{\vec{n}_s}s(\vec{n}_s,n_t)
\rho^{a^\dagger,a}(\vec{n}_s)\right]:.
\eeq
In this case there are no sign-oscillations for even number of nucleons~\cite{Chen:2004rq}
and we do have only one auxiliary
field such that the simulations are much cheaper. Although there is no
positivity theorem for odd numbers of nucleons, sign oscillations appear to be
suppressed also in systems with odd number of nucleons because it is only one 
particle away from an even system with no sign-oscillation. 
Since the final result is closed to the one
 produced by ${\rm SU}(4)$-symmetric simulation we
divide our simulations in three parts. To simulate an expectation value of
some observable we use ${\rm SU}(4)$-symmetric transfer matrices in the first
and the last $L_{t_0}$ steps in order to filter the low-energy signal and after filtering
start the simulation with realistic transfer matrices. A schematic overview of
the transfer matrix calculation is shown in Fig.~\ref{timesteps}.  

For our nuclear lattice simulations we use the hybrid Monte-Carlo (HMC)
method~\cite{Duane:1987de}. We introduce the conjugate fields $p_{\pi_I}$,
$p_s$, $p_{s_I}$ and use molecular dynamics trajectories to generate new 
configurations for the fields $p_{\pi_I}$, $p_s$, $p_{s_I}$, $\pi_I$, $s$,
$s_I$ which keep the HMC Hamiltonian
\beq
H_{{\rm HMC}}=\frac{1}{2}\sum_{\vec{n}}\left(\sum_I\left[p_{\pi_I}^2(\vec{n})+p_{s_I}^2(\vec{n})\right]+p_s^2(\vec{n})\right)+V(\pi_I,s,s_I),
\eeq
constant, where the HMC potential is defined by
\beq
V(\pi_I,s,s_I)=S_{\pi\pi}+S_{ss}-\log|\det {\cal M}|.
\eeq
Upon completion of each molecular dynamics trajectory, we apply Metropolis 
accept or reject step for the new configuration according to the probability
distribution $\exp(-H_{{\rm HMC}})$. This process of molecular dynamics 
trajectory and Metropolis step is repeated many times.

\section{Leading-order results}

\begin{figure}[t]
\vspace{-1.5cm}
\begin{center}
\includegraphics[width=10.0cm,keepaspectratio,angle=0,clip]{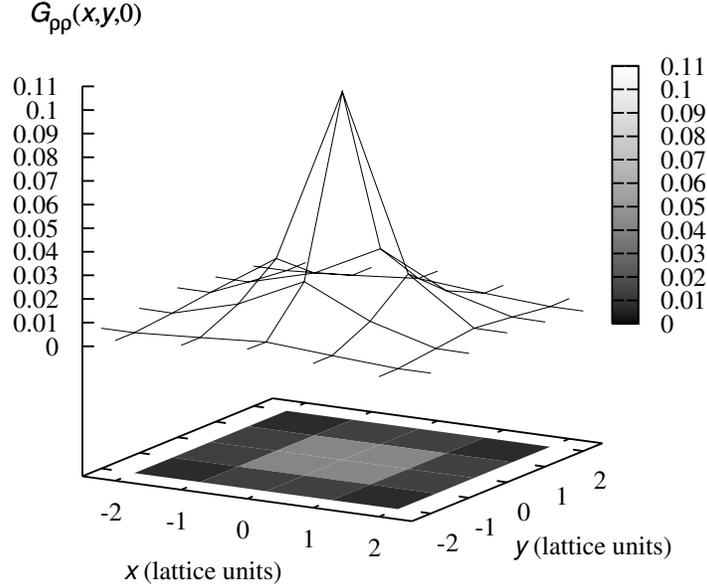}
\vspace{-1cm}
\caption[tritoncorr]{\label{tritoncorr} The nucleon density correlation for
  the triton in $xy$-plane.}
\vspace{0.2cm}
\end{center}
\end{figure}

With the presented method we performed nuclear lattice simulations on
JUBL/JUGENE supercomputer at Forschungszentrum J\"ulich. Already at leading 
order we get promising results for binding energies, radii and density
correlations for the deuteron, triton and helium-4~\cite{Borasoy:2006qn}. 
Numerical results on a 
$5^3$ lattice for triton and helium-4 are shown in Table~\ref{tab:2}. The
triton binding energy agrees with experiment within $5\%$ and the triton
root-mean-square radius is accurate to $30\%$. The binding energy for
helium-4 is within $25\%$ of the experimental value while the root-mean-square 
radius agrees within $10\%$. Our results for the triton nucleon density 
correlations are shown in Fig.~\ref{tritoncorr}. We also studied the feasibility of 
simulations for light nuclei with up to eight nucleons and observed that for
$A\le 8$ the CPU time scales approximately linear with A.

\begin{table*}[t] 
\begin{center}
\begin{tabular*}{0.7\textwidth}{@{\extracolsep{\fill}}||l||c|c|c|c||}
    \hline \hline
{} & {} &  {} &  {} & {}\\[-1.5ex]
    & {$E_{{\rm 3H}}[{\rm MeV}]$}    & {$r_{{\rm 3H}}[{\rm fm}]$}  & {$E_{{\rm 4He}}[{\rm MeV}]$}   &  {$r_{{\rm 4He}}[{\rm fm}]$} \\[1ex]
\hline  \hline
{} & {} &  {} &  {} & {}\\[-1.5ex]
{Simulation} &   {$-8.9(2)$}   & {$2.27(7)$}   & {$-21.5(9)$}  & {$1.50(14)$} \\[0.3ex]
{Experiment} &  {$-8.482$}    & {$1.755(9)$}   & {$-28.296$}     & {$1.673(1)$} \\[1ex]
    \hline \hline
  \end{tabular*}
\caption{Experimental and nuclear lattice simulation results for binding energies and root-mean-square radius of triton and helium-4.  
    \label{tab:2}}
\end{center}
\end{table*}

\section{Next-to-leading-order results}

At NLO there appear 9 low energy constants (LECs) which we fitted to the
Nijmegen NN scattering data. Elastic scattering phase-shifts on the lattice 
are related by L\"uscher's~\cite{Luscher:1985dn,Luscher:1986pf,Luscher:1990ux} formula to the energy levels of two-body
states in a finite large volume cubic box with periodic boundary
conditions. While this method is very useful at low momenta, it is not so
useful for determining phase shifts on the lattice at higher energies and
higher orbital angular momenta. Furthermore, spin-orbit coupling and
partial-wave mixing are difficult to measure accurately using L\"uscher's
method due to multiple-scattering artifacts produced by the periodic cubic
boundary conditions. In~\cite{Borasoy:2007vy} we proposed a more robust
approach to measure phase shifts for two nonrelativistic point particles
on the lattice using a spherical wall boundary. The basic idea is to impose
a hard spherical wall boundary on the relative separation between the two 
interacting particles at some chosen radius. The reason for
this spherical wall is to remove copies of the two-particle interactions
due to the periodic boundaries on the lattice. This additional boundary condition 
allows for a direct extraction of the phase-shifts and mixing angles from the 
finite-volume spectrum. For more details see~\cite{Borasoy:2007vy}.

\begin{figure}[t]
\vspace{-1.5cm}
\begin{center}
\includegraphics[width=15.0cm,keepaspectratio,angle=0,clip]{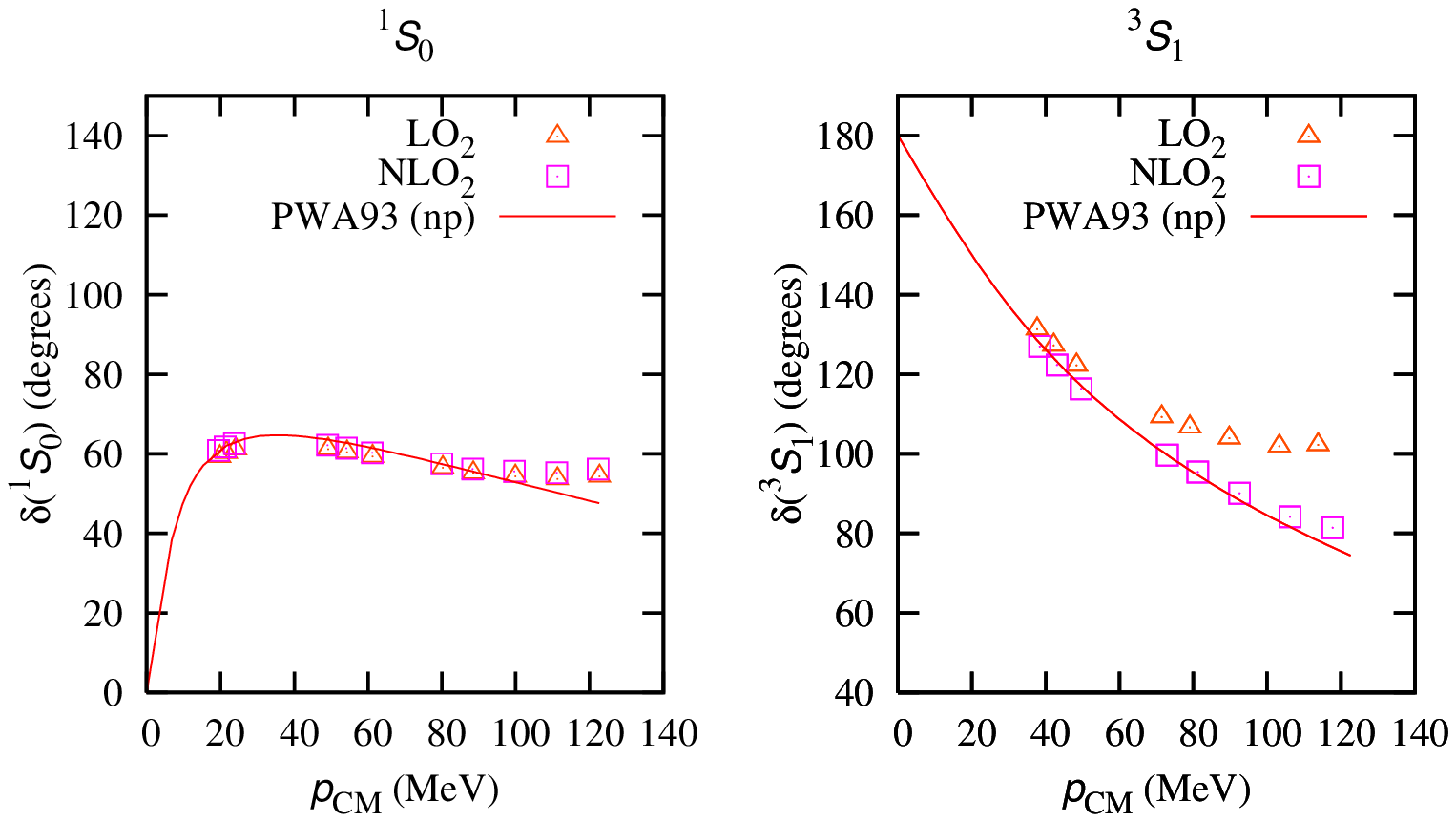}
\vskip -1.5 true cm
\includegraphics[width=15.0cm,keepaspectratio,angle=0,clip]{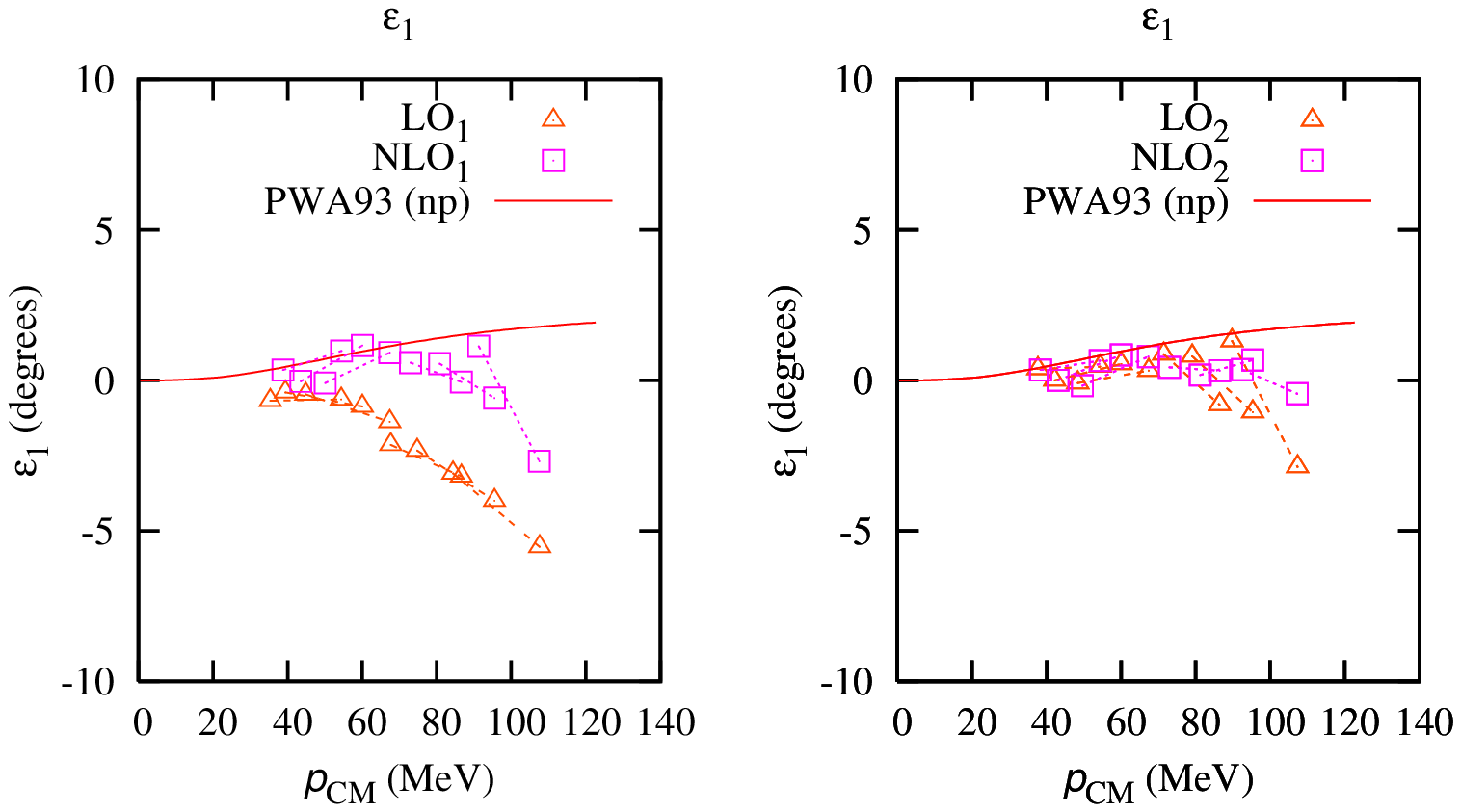}
\vspace{-0.8cm}
\caption[nlophaseshifts]{\label{nlophaseshifts} NN S-wave phase
  shifts and mixing angles versus
  center-of-mass momentum with actions ${\rm LO}_1$ and ${\rm LO}_2$.}
\end{center}
\end{figure}

Using the spherical wall method we determined the values of 9 LECs by matching
three S-wave, four P-wave scattering data points, as well as deuteron binding
energy and quadrupole moment. In Fig.~\ref{nlophaseshifts}  we show NN S-wave phase-shifts and
mixing angles for two different actions, called ${\rm LO}_1$ and ${\rm LO}_2$. The action
${\rm LO}_1$ is the one presented in Eq.~\ref{trasferMLO1}. In the action 
${\rm LO}_2$ the contact interactions are smeared by a Gaussian. The two
actions are identical at  leading order and differ only by higher-order
terms. The main motivation to introduction the Gaussian smearing was to cure a multi-particle
clustering instability at coarse lattice spacing present in simulations with
${\rm LO}_1$ and to estimate a systematic error coming from higher-order
corrections see~\cite{Borasoy:2006qn} for extended discussion.
As can be seen from Fig.~\ref{nlophaseshifts}, our lattice simulation results
are in a good agreement with the partial wave results for momenta smaller than
$80$~MeV. Deviations between the two results for different actions appear
merely at larger momenta and are consistent with the expected higher order effects.

\section{Dilute neutron matter}
As a first application at NLO we simulate dilute neutron matter
in a periodic box~\cite{Borasoy:2007vk}. We probe the density range from 
$2\%$ to $8\%$ of normal
nuclear matter density. Neutron-rich matter at this density is likely to be present
in the inner crust of neutron stars. The Pauli suppression of three-body
forces in dilute neutron matter makes it a good testing ground for chiral
EFT applied to many-nucleon systems. Neutron matter
at $k_F\sim 80$~MeV, where 
\beq
k_F=\frac{1}{L}(3\pi^2 N)^{1/3}
\eeq
is Fermi momentum, is close to the so-called idealized unitary limit. In this
limit the S-wave scattering length is infinite and the range of the
interaction is zero such that the scattering amplitude is as strong as 
possible. At lower densities corrections due to the finite scattering length
become more important while at higher densities corrections due to 
effective range start to dominate.
In the unitary limit the ground state has no dimensionful parameters
other than the particle density and so the ground state energy of the system
should obey a simple relation
\beq
E_0=\xi E_0^{{\rm free}}
\eeq
where $\xi$ is a dimensionless measurable constant. Due to its universal nature, 
the unitary limit can be studied in atomic systems. Ultracold $^6{\rm Li}$
and $^{40}{\rm K}$ atoms e.g. can be tuned into the unitary limit by using
a magnetic-field Feshbach resonance. Recently measured values for $\xi$
scatter considerably and have large error bars:
\beq
\xi=0.51(4)\cite{Kinast0502},0.46_{-05}^{+12}\cite{Stewart0607},0.32_{-13}^{+10}\cite{Bartenstein02}.
\eeq
Earlier experiments tend to yield larger value for $\xi$ indicating the need
of further experimental studies.

\begin{figure}[t]
\begin{center}
\includegraphics[width=8.0cm,keepaspectratio,angle=0,clip]{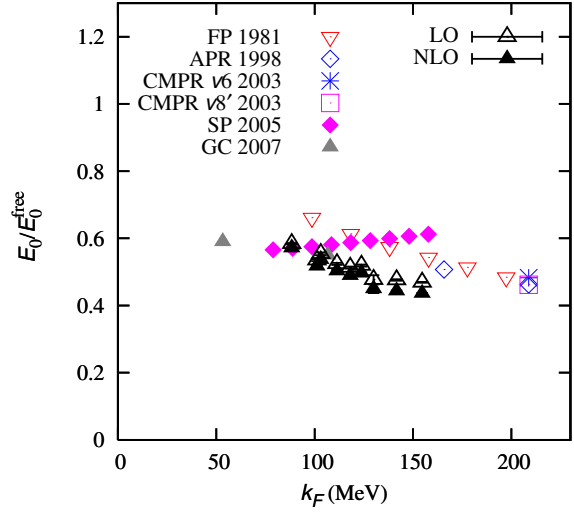}
\vspace{-0.15cm}
\caption[xiliter]{\label{xiliter} Results for $E_0/E_0^{{\rm free}}$ versus
  Fermi momentum $k_F$. For comparison we show the results for 
FP~1981~\cite{Friedman:1981qw}, APR~1998~\cite{Akmal:1998cf}, 
CMPR~$v6$ and $v8^\prime$~2003~\cite{Carlson:2003wm},
SP~2005~\cite{Schwenk:2005ka} and GC~2007~\cite{Gezerlis:2007fs}}
\end{center}
\vspace{0.2cm}
\end{figure}

There have been numerous analytic calculations of $\xi$,
see~\cite{Furnstahl:2008df} for a recent review. The obtained values for $\xi$ vary
roughly from $0.2$ to $0.6$. To get a nuclear lattice EFT picture of the neutron
matter in the unitary regime we simulate the ground state of $8,12$ and $16$ 
neutrons in a box of length $L=10,12$ and $14$~fm using Monte Carlo. In
Fig.~\ref{xiliter} we show ground energy ratio $E_0/E_0^{{\rm free}}$ 
in dependence
of Fermi momentum $k_F$. For comparison we also show earlier phenomenological
calculations. Our predictions seem to be consistent with the earlier 
results. We find a good fit to the lattice data using
\beq
E_0/E_0^{{\rm free}}\simeq \xi-\frac{\xi_1}{k_F\,a_{{\rm scatt}}}+
0.16\,\, k_F\,r_{{\rm eff}}-(0.51{\rm fm}^3) k_F^3.
\eeq
The results from the fit are
\beq
\xi\simeq 0.31\quad{\rm and}\quad\xi_1\simeq 0.81.
\eeq

\section{N$^2$LO three-body forces}
\begin{figure}[t]
\begin{center}
\includegraphics[width=9.2cm,keepaspectratio,angle=0,clip]{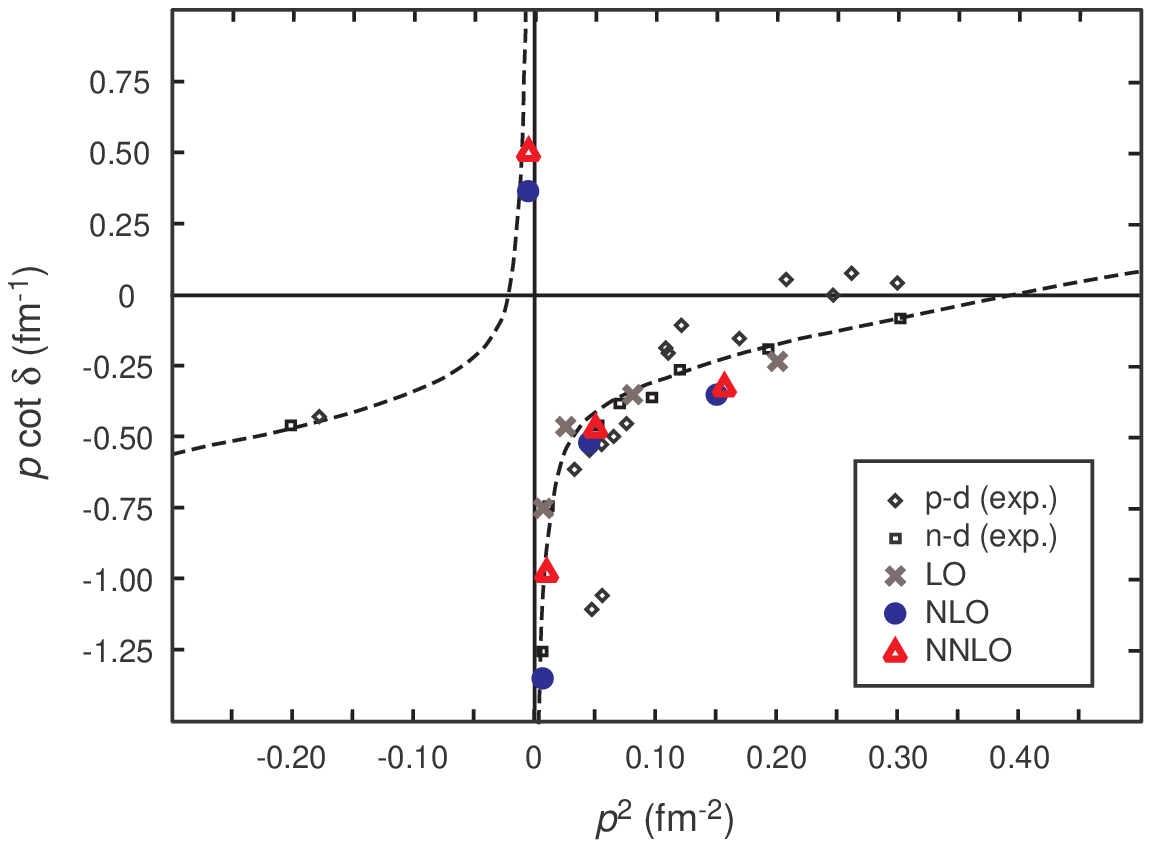}~\includegraphics[width=7.0cm,keepaspectratio,angle=0,clip]{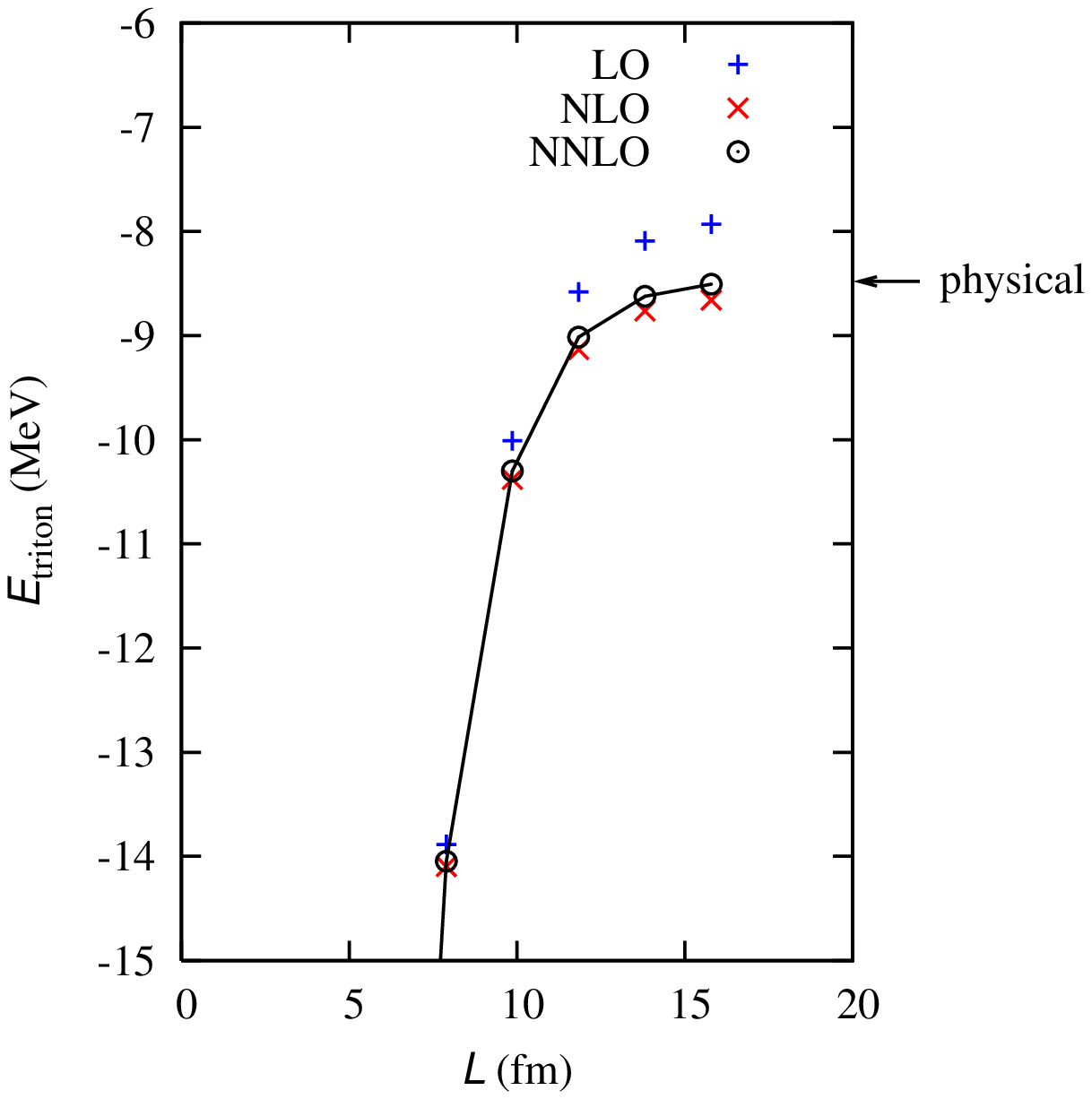}
\vspace{-0.15cm}
\caption[nddoublet]{\label{nddoublet} In the left panel: 
neutron-deuteron S-wave scattering phase-shifts in the spin-$1/2$ doublet channel versus 
the square of relative momentum. For completeness we show experimental 
data~\cite{oers67nd} for 
proton-deuteron and neutron-deuteron scattering.
In the right panel: triton binding energy
versus the length of the lattice box.}
\end{center}
\vspace{0.2cm}
\end{figure}

\begin{figure}[t]
\begin{center}
\includegraphics[width=9.2cm,keepaspectratio,angle=0,clip]{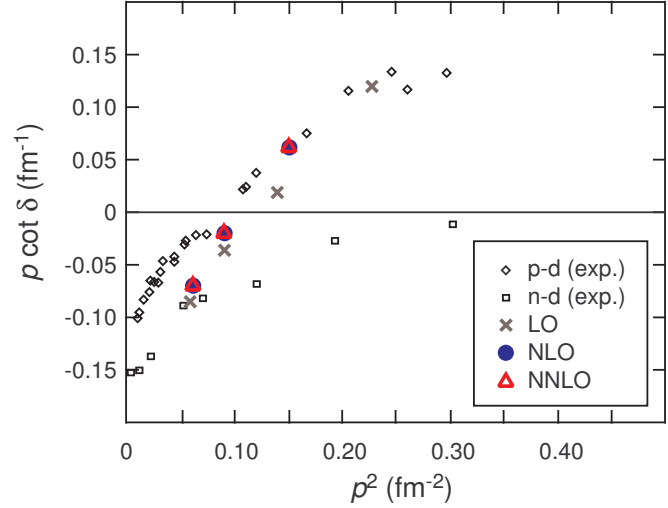}
\vspace{-0.15cm}
\caption[ndquartet]{\label{ndquartet} 
Neutron-deuteron scattering S-wave phase-shifts in the spin-$3/2$ quartet channel versus 
the square of relative momentum. The
data for 
proton-deuteron and neutron-deuteron scattering are taken from~\cite{oers67nd}.
}
\end{center}
\end{figure}

At N$^2$LO three-body forces start to show up which depend on two constants. 
We fit these LECs from neutron-deuteron scattering data in the
spin-$1/2$ doublet channel and the triton binding energy. To describe the
neutron-deuteron scattering we use standard L\"uscher 
formula~\cite{Luscher:1985dn,Luscher:1986pf,Luscher:1990ux}. Finite 
volume spectrum was generated with Lanczos diagonalization method.
In Fig.~\ref{nddoublet} we show the S-wave phase-shift in spin-$1/2$ doublet channel 
versus the square of the relative momentum and the triton binding energy versus the 
length of the lattice box. One observes a very natural convergence pattern in our 
simulations with increasing chiral order. Probing the triton binding energy
we see from Fig.~\ref{nddoublet} that at the box length $\sim15$~fm the 
volume dependence already becomes very small and the binding energy approaches its 
physical value. This is consistent with our expectation that the volume
dependence in nuclear lattice EFT simulations should become weak for 
$L\sim 20$~fm. In Fig.~\ref{ndquartet} we show the S-wave phase-shifts in the
spin-$3/2$ quartet 
channel
versus the square of relative momentum. This channel was not taken into 
account in the fit procedure. Again we observe a very nice convergence with
increasing chiral order. Our predictions are located between the proton-deuteron
and neutron-deuteron experimental data. Since the isospin-breaking was
not taken into account in our simulations the results are very satisfactory.  

\begin{figure}[t]
\begin{center}
\includegraphics[width=9.2cm,keepaspectratio,angle=0,clip]{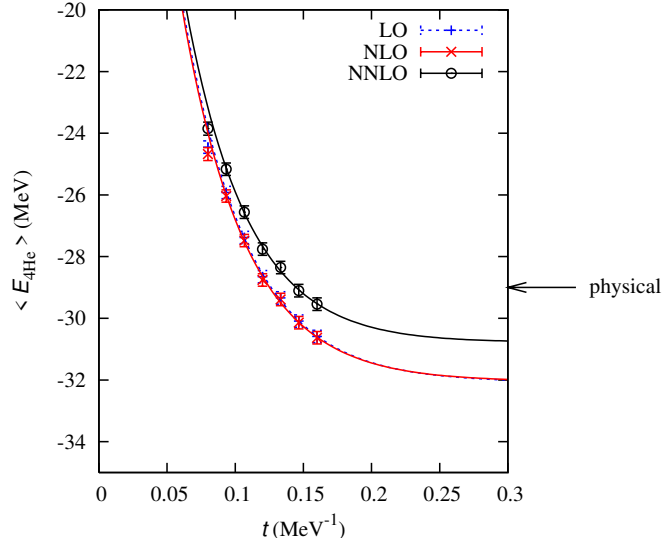}
\vspace{-0.15cm}
\caption[bindingEnHe4]{\label{bindingEnHe4} 
Binding energy expectation value of $^4{\rm He}$ versus Euclidean time $t$. Plot
produced by Monte-Carlo simulation with the box length $L=16$~fm.}
\end{center}
\end{figure}

As a first Monte-Carlo simulation  of N$^2$LO lattice EFT we studied the binding
energy of $^4{\rm He}$. The length of the box was chosen $L=16$~fm. 
In Fig.~\ref{bindingEnHe4} we show the resulting binding energy of the $^4{\rm He}$ system
\beq
 \langle E_{4{\rm He}}\rangle =
\frac{\langle\Psi_4|\exp(-t\,H/2)H\exp(-t\,H/2)|\Psi_4\rangle}{
\langle\Psi_4|\exp(-t\,H)|\Psi_4\rangle}
\eeq
versus Euclidean time $t$. Our Monte-Carlo simulations overpredict  the
physical binding energy with subtracted Coulomb-effects by $5\%$. 
This is consistent with the expected theoretical accuracy of our simulations.

\section{Summary and outlook}
The results of our studies demonstrate that lattice EFT is a promising tool for a quantitative
description of light nuclei. At leading order binding energies and radii
of nuclei up to $^4{\rm He}$ are reproduced with the accuracy  
$5\dots 30\%$. At NLO, $9$ LECs were fitted to the NN scattering
phase-shift using the spherical wall method which is best suited  
 to measure phase shifts and mixing angles for nonrelativistic point particles
on the lattice. With the NLO EFT action, we studied dilute neutron matter close
to the unitary limit. We performed Monte-Carlo simulation with $N=8,12$ and $16$
neutrons in a box of length $L=10,12$ and $14$~fm. Our simulation results
are consistent with earlier phenomenological determinations. We also
presented the first analysis of N$^2$LO lattice EFT. At this order, the two LECs
entering the
 three-body
force were fitted to neutron-deuteron scattering data and the triton binding
energy. In our first N$^2$LO Monte-Carlo simulation we calculated the binding
energy of $^4{\rm He}$. Our simulations overpredict the physical binding energy
of  $^4{\rm He}$ by $\sim 5\%$ which is within the expected accuracy of our
lattice simulations. 

In the future, we plan to perform N$^2$LO Monte-Carlo simulations of light 
nuclei and probe neutron matter with larger number of neutrons in a box.

\vspace{0.2cm}



\end{document}